\documentclass[12pt]{article}
\usepackage{amsmath}
\usepackage{amssymb}
\usepackage{graphicx}
\usepackage{epsfig}
\usepackage{subfigure}
\usepackage{cite}
\usepackage{url}
\usepackage[small]{caption}
\begin{document}
\baselineskip 0.6cm

\def\simgt{\mathrel{\lower2.5pt\vbox{\lineskip=0pt\baselineskip=0pt
           \hbox{$>$}\hbox{$\sim$}}}}
\def\simlt{\mathrel{\lower2.5pt\vbox{\lineskip=0pt\baselineskip=0pt
           \hbox{$<$}\hbox{$\sim$}}}}
\def\simprop{\mathrel{\lower3.0pt\vbox{\lineskip=1.0pt\baselineskip=0pt
             \hbox{$\propto$}\hbox{$\sim$}}}}
\def\bra#1{\langle #1 |}
\def\ket#1{| #1 \rangle}
\def\inner#1#2{\langle #1 | #2 \rangle}

\begin{titlepage}

\begin{flushright}
UCB-PTH-13/10\\
\end{flushright}

\vskip 1.5cm

\begin{center}

{\Large \bf
Grand Unification \\ \vspace{-0.1cm}
and \\ \vspace{0.1cm}
Intermediate Scale Supersymmetry
}

\vskip 0.8cm

{\large Lawrence J. Hall and Yasunori Nomura}

\vskip 0.4cm

{\it Berkeley Center for Theoretical Physics, Department of Physics,\\
 University of California, Berkeley, CA 94720, USA}

\vskip 0.1cm

{\it Theoretical Physics Group, Lawrence Berkeley National Laboratory,
 CA 94720, USA}

\vskip 0.8cm

\abstract{With minimal field content and for an interesting range of the 
supersymmetric Higgs mixing parameter, $0.5 \simlt \tan^2\!\beta \simlt 2$, 
the superpartner mass scale, $\tilde{m}$, is found to be at the intermediate 
scale, $\sim 10^{10 \pm 1}~{\rm GeV}$, near where the Standard Model Higgs 
quartic coupling passes through zero.  For any 4d supersymmetric grand 
unified symmetry spontaneously broken by a vacuum expectation value 
$\langle \Sigma \rangle$, if superpotential interactions for $\Sigma$ 
are forbidden e.g.\ by $R$ symmetries, the uneaten color octet, $\Sigma_8$, 
and weak triplet, $\Sigma_3$, have masses of order $\tilde{m}$.  The 
combination of superpartner and $\Sigma_{8,3}$ states leads to successful 
gauge coupling unification, removing the disastrously high proton decay 
rate of minimal Standard Model unification.  Proton decay could be seen 
in future experiments if $\tilde{m} \sim 10^{11}~{\rm GeV}$, but not if 
it is lower.  If the reheating temperature after inflation, $T_R$, is less 
than $\tilde{m}$ dark matter may be axions.  If $T_R > \tilde{m}$, thermal 
LSP dark matter may lead to the environmental selection of a TeV-scale LSP,  
either wino or Higgsino, which could comprise all or just one component 
of dark matter.  In the Higgsino case, the dark matter is found to behave 
inelastically in direct detection experiments, and gauge coupling unification 
occurs accurately without the need of any threshold corrections.}

\end{center}
\end{titlepage}

\section{Introduction}
\label{sec:intro}

The discovery of a Higgs boson with mass near $125~{\rm GeV}$ not only 
completes the Standard Model (SM), but allows the SM to provide a perturbative 
description of nature up to extremely high energies.  Such a ``desert'' 
hypothesis was first suggested in the context of the unification of the 
SM into an $SU(5)$ grand unified theory~\cite{Georgi:1974sy,Georgi:1974yf}. 
In the following four decades, data consistently kept open this possibility, 
even while the theoretical emphasis was on extended models of particle 
physics at the weak scale. Precision measurements of gauge couplings were 
broadly consistent with their perturbative unification at very high energies, 
neutrino masses were consistent with dimension-5 operators suppressed by 
a large mass near the unified scale, and all quark and charged lepton masses 
were found to be sufficiently light that the corresponding Yukawa couplings 
remained perturbative to unified energies.  A Higgs boson lighter than 
about $110~{\rm GeV}$ would exclude such a large energy desert, as the 
electroweak vacuum would be unstable with a lifetime much less than 
$10^{10}~{\rm years}$; and a Higgs boson heavier than about $190~{\rm GeV}$ 
would have led to the Higgs quartic coupling becoming non-perturbative 
well below the unified mass, heralding a new scale of non-perturbative 
physics.  Perhaps even more remarkable than the discovery of a Higgs boson 
in this high energy desert window of $110~\mbox{--}~190~{\rm GeV}$, is 
the absence of a single discovery of physics beyond the SM during these 
four decades:\ an extension of the SM at the weak scale has had ample 
opportunity to reveal itself in rare flavor and $CP$ violating processes, 
at a succession of collider experiments, culminating in the negative 
results at the LHC so far, and in both direct and indirect searches for 
dark matter in our galaxy.

The most glaring shortcoming of the desert hypothesis is that the weak 
scale is highly unnatural, requiring many orders of magnitude of fine-tuning. 
Indeed, this was the principle driver behind theoretical efforts at 
extending the SM at the weak scale and experimental programs for beyond 
SM physics at LEP, Tevatron, and the LHC.  Yet, as with the cosmological 
constant~\cite{Weinberg:1987dv}, this fine-tuning may result from 
environmental selection on a multiverse.  While the physical origin 
of this selection is uncertain, one possibility is the requirement 
of complex nuclei~\cite{Agrawal:1997gf}.

While experiment indicates a broad picture of perturbative unification 
of the SM, there are difficulties with the simplest implementation.  First 
and foremost, the precision of gauge coupling unification is limited:\ a 
$6\%$ correction at the unified scale is required and, furthermore, to be 
consistent with limits on proton stability this correction must raise the 
superheavy gauge boson mass $M_X$ from about $10^{14}~{\rm GeV}$ to above 
$6 \times 10^{15}~{\rm GeV}$.  Secondly, the mass ratios of down-type quarks 
to charged leptons do not agree with the simplest $SU(5)$ expectations.

In this paper we seek an origin for precision gauge coupling unification 
that is linked to the value of the Higgs boson mass.  The Higgs quartic 
coupling, when evolved to high energies, becomes negative and is 
small.  Is there a simple physical understanding of this behavior? 
One possibility is that the quartic scans in the landscape with a 
distribution favoring negative values, and that our universe is typical 
among those with a sufficiently long-lived electroweak vacuum to allow 
observers~\cite{Feldstein:2006ce}.  In this paper we pursue an alternative 
idea---the underlying theory of nature is supersymmetric and the scale 
of the SM superpartners, $\tilde{m}$, is close to the scale at which 
the Higgs quartic coupling $\lambda$ passes through zero:
\begin{equation}
  \tilde{m} \sim \mu_c
\quad\mbox{with}\quad
  \lambda(\mu_c) = 0.
\label{eq:quartic-bc}
\end{equation}
Since the current experimental data indicates $\mu_c \simeq 
10^{10}~\mbox{--}~10^{12}~{\rm GeV}$~\cite{Buttazzo:2013uya}, this 
leads to the picture of Intermediate Scale Supersymmetry (ISS).

Values of $\tilde{m}$ as high as the unified scale are possible if one takes 
into account the uncertainty of the top quark mass determination at a level 
of $2~\mbox{--}~3\sigma$.  Indeed, we previously predicted that in High 
Scale Supersymmetry, with the SM valid up to unified scales where it becomes 
supersymmetric, the Higgs boson mass is $(128 \pm 3)~{\rm GeV}$ for $\tan\beta 
= 1$~\cite{Hall:2009nd}; see also~\cite{Hebecker:2012qp,Ibanez:2013gf} 
for more recent analyses.  This prediction was found to be highly 
insensitive to the scale of $\tilde{m}$, dropping by only $1~{\rm GeV}$ 
as the scale is reduced from $10^{14}~{\rm GeV}$ to $10^{12}~{\rm GeV}$. 
Furthermore, $\tan\beta$ of unity can result from an approximate discrete 
symmetry that interchanges the two Higgs doublets.

Here we take the view that the supersymmetric boundary condition in 
Eq.~(\ref{eq:quartic-bc}) occurs substantially below the unified scale, 
allowing us to connect predictions for the Higgs boson mass with those 
for precision gauge coupling unification and proton decay.  Thus we 
assume that the unified symmetry is broken at scale $M_X$ to the Minimal 
Supersymmetric Standard Model (MSSM), with supersymmetry broken at 
$\tilde{m} \ll M_X$.  In any such unified scheme, the superheavy gauge 
particles acquire mass by eating part of an $SU(5)$ adjoint multiplet 
$\Sigma$.  The remaining states of $\Sigma$, with mass $M_\Sigma$, are 
therefore in a $SU(5)$-split representation and contribute to gauge 
coupling unification.  One frequently assumes that $M_\Sigma \sim M_X$, 
but this need not be the case.  A symmetry may forbid a $\Sigma$ mass 
term from appearing in the superpotential, while allowing it to arise 
from supersymmetry breaking, so that $M_\Sigma \sim \tilde{m}$.  This is 
very much analogous to $\mu \sim \tilde{m}$ in the MSSM.  The combination 
of gauginos and $\Sigma$ improves gauge coupling unification, and allows 
$M_X > 6 \times 10^{15}~{\rm GeV}$ for $\tilde{m} \simlt 2 \times 
10^{11}~{\rm GeV}$, making it consistent with the constraints from 
proton decay~\cite{Nishino:2012ipa} without resorting to a special 
mechanism at the unified scale.

The ISS framework discussed in this paper allows for three different 
manifestations, depending on how the environmental selection effect works 
in the multiverse.  The first, in some sense the simplest, possibility 
is that the masses of all the superpartners are within a couple of orders 
of magnitude at $\tilde{m}$.  The other two cases are that the wino and 
Higgsino masses, respectively, are brought down to the TeV scale so that 
they comprise (a part of) the dark matter; all the other superpartner 
masses stay around $\tilde{m}$.  In the case of Higgsino dark matter, 
splitting between the two neutral components are of $O(100~{\rm keV})$ 
naturally yielding inelastic dark matter~\cite{TuckerSmith:2001hy}. 
Precision gauge coupling unification may be analyzed in each case.  In 
the case of TeV-scale Higgsinos the ratios of the observed gauge couplings 
are reproduced from the unification condition at the level of the MSSM 
without extra threshold corrections, while in the other two cases threshold 
corrections from a dimension-5 operator, of natural size, are required 
at the unified scale.  In either of the three cases, however, the mass 
of the superheavy gauge bosons can be determined with some precision, 
giving relatively tight implications for the proton decay rate.

The organization of the rest of the paper is as follows. 
In Section~\ref{sec:higgsmass}, we discuss the condition in 
Eq.~(\ref{eq:quartic-bc}), motivating ISS.  In Section~\ref{sec:model}, 
we introduce our minimal unified theory with ISS in which $M_\Sigma \sim 
\tilde{m}$ arises in a simple manner.  In Section~\ref{sec:multiverse}, 
we discuss supersymmetry breaking and environmental selection for dark 
matter.  We consider three different manifestations of our framework:\ 
pure ISS, ISS with TeV-scale wino, and ISS with TeV-scale Higgsino. 
Gauge coupling unification and proton decay are discussed in each case.

\section{{\boldmath $\tilde{m}$} from the Zero of the Higgs Quartic}
\label{sec:higgsmass}

We take the theory below the unified scale to be the MSSM with the mass 
squared matrix for the two Higgs doublets given by
\begin{equation}
  {\cal M}_{\rm H}^2 = \begin{pmatrix}
    |\mu|^2 + m_{H_u}^2 & \mu B \\
    \mu B & |\mu|^2 + m_{H_d}^2
 \end{pmatrix},
\label{eq:higgsmasssquared}
\end{equation}
where the soft supersymmetry breaking parameters $m_{H_u,H_d}^2$ and $B$ 
are governed by the scale $\tilde{m}$.  As we will see later, we consider 
that the unified theory above the unified scale is the minimal supersymmetric 
$SU(5)$ model~\cite{Dimopoulos:1981zb} or its simple extensions, e.g.\ 
to $SO(10)$, so that the supersymmetric mass parameter $\mu$ arises from 
a cancellation of terms of order the unified scale.  Note that such a 
cancellation occurs as a consequence of environmental selection of having 
the correct electroweak symmetry breaking, which requires $|\mu| \simlt 
\tilde{m}$.

The theory below $\tilde{m}$ is the SM, possibly with a few supersymmetric 
partners required by other environmental requirements.  (We will later 
consider the case in which the Higgsinos or winos are at a TeV scale due 
to environmental selection associated with dark matter.)  This requires 
the determinant of the matrix ${\cal M}_{\rm H}^2$ to be extremely 
small compared with its natural size $\sim \tilde{m}^4$, forced by the 
environmental condition of electroweak symmetry breaking.  Writing the 
SM Higgs potential as
\begin{equation}
  V(h) = -m^2 h^\dagger h + \frac{\lambda}{4} (h^\dagger h)^2,
\label{eq:higgspot}
\end{equation}
the quartic coupling at the scale $\tilde{m}$ is dominated by the tree-level 
result
\begin{equation}
  \lambda(\tilde{m}) = \frac{g^2 + g'^2}{2} \, \cos^2\! 2\beta
\qquad\mbox{with}\qquad
  \tan^2\!\beta = \frac{|\mu|^2 + m_{H_d}^2}{|\mu|^2 + m_{H_u}^2},
\label{eq:lambdamtilde}
\end{equation}
where the $SU(2) \times U(1)$ gauge couplings $g, g'$ and the parameters 
$\mu, m_{H_u}^2, m_{H_d}^2$ are all evaluated at scale $\tilde{m}$.

A key point is that there is a large region of parameter space in which 
$\lambda(\tilde{m})$ is very small, so that Eq.~(\ref{eq:quartic-bc}) 
can be used to obtain the correct order of magnitude of $\tilde{m}$.  To 
begin with, suppose that an approximate symmetry leads to $m_{H_u,H_d}^2$ 
being approximately equal, allowing an expansion $\tan^2\!\beta = 1 + 
\epsilon$, with $\epsilon \ll 1$.  One discovers that $\lambda(\tilde{m})$ 
is remarkably small, arising only at quadratic order
\begin{equation}
  \lambda(\tilde{m}) \simeq 0.06\, \epsilon^2 + O(\epsilon^3),
\label{eq:lambdaexpansion}
\end{equation}
where the coefficient reflects the values of the gauge couplings at 
the scale $\sim 10^{10}~{\rm GeV}$, where the SM Higgs quartic is near 
zero.  Indeed, in the SM the Higgs quartic coupling vanishes at $\mu_c 
\simeq 10^{10}~\mbox{--}~10^{12}~{\rm GeV}$, but providing $\lambda(\tilde{m}) 
< 0.02$, $\tilde{m}$ is less than an order of magnitude below $\mu_c$. 
Comparing with Eq.~(\ref{eq:lambdaexpansion}) we see that $\epsilon$ need 
not be very small for this to happen.  Thus we can take the condition 
of vanishing $\lambda$ to determine the correct order of magnitude for 
$\tilde{m}$ provided $\lambda(\tilde{m}) < 0.02$, which translates into 
\begin{equation}
  0.55 \; \simlt \; \tan^2\!\beta 
  = \frac{|\mu|^2 + m_{H_d}^2}{|\mu|^2 + m_{H_u}^2} \; \simlt \; 1.8,
\label{eq:t^2range}
\end{equation}
where we have used the full expression in Eq.~(\ref{eq:lambdamtilde}), 
instead of Eq.~(\ref{eq:lambdaexpansion}).  This does not require near 
degeneracy between $m_{H_u,H_d}^2$; indeed they can differ by much more 
than is generated by renormalization group scaling from the unified 
scale using the top Yukawa coupling.

\section{Minimal Unified Theory with ISS}
\label{sec:model}

The simple group of lowest rank that allows unification of 
the known gauge forces is $SU(5)$~\cite{Georgi:1974sy}.  With 
supersymmetry restored at the intermediate scale, $\tilde{m} 
\sim \mu_c \simeq 10^{10}~\mbox{--}~10^{12}~{\rm GeV}$, we seek 
a supersymmetric unified theory, and the minimal field content 
is well known~\cite{Dimopoulos:1981zb}:\ Higgs chiral superfields 
$\Sigma({\bf 24})$ to break $SU(5)$ and $H({\bf 5}), \bar{H}({\bf 5}^*)$ 
to break the electroweak interaction, together with matter chiral superfields 
$T({\bf 10})$ and $\bar{F}({\bf 5}^*)$ containing quarks and leptons. 
Below, we assume that $R$ parity is conserved.  We omit the generation 
index for quarks and leptons throughout.

There are four possible interactions in the superpotential involving 
the Higgs multiplets that have dimension 4 or less:\ two mass terms, 
$[\bar{H} H]_{\theta^2}$ and $[\Sigma^2]_{\theta^2}$, and two trilinear 
interactions, $[H \Sigma \bar{H}]_{\theta^2}$ and $[\Sigma^3]_{\theta^2}$. 
This should be compared to the MSSM, where there is a single such possible 
Higgs interaction in the superpotential $[\mu H_u H_d]_{\theta^2}$.  The 
phenomenology of weak scale supersymmetry requires that $\mu \sim \tilde{m} 
\sim {\rm TeV}$, the well known ``$\mu$'' problem of the MSSM.  With 
high scale mediation of supersymmetry breaking, it is easy to solve this 
problem by forbidding such a superpotential term and instead allowing 
a bilinear interaction between the two Higgs field in the K\"{a}hler 
potential, $[H_u H_d]_{\theta^4}$, since in supergravity this leads to 
an effective term in the superpotential with coefficient proportional 
to the gravitino mass, $\mu = c m_{3/2}$, where $c$ is a constant of 
$O(1)$~\cite{Giudice:1988yz}.

With ISS we can ask whether either of the Higgs mass terms, $H \bar{H}$ 
and $\Sigma^2$, might similarly appear in the K\"{a}hler potential rather 
than in the superpotential.  For $H \bar{H}$ we choose to put it in the 
superpotential, which makes the colored triplets in these multiplets have 
a mass of order the unified scale. The superpotential of the minimal ISS 
theory then contains
\begin{equation}
  W = m_H H \bar{H} + \lambda_H H \Sigma \bar{H} 
    + y_U TTH + y_D T \bar{F} \bar{H}.
\label{eq:W}
\end{equation}
If the theory possesses an $R$ symmetry in the supersymmetric limit, then 
the presence of the first two terms forbids $\Sigma^2$ and $\Sigma^3$ 
from appearing in the superpotential $W$, while allowing them in the 
K\"{a}hler potential
\begin{equation}
  K \,\supset\, \frac{c_2}{2} \Sigma^2 + \frac{c_3}{3 \Lambda} \Sigma^3,
\label{eq:K}
\end{equation}
where $c_{2,3}$ are dimensionless couplings of order unity, while $\Lambda$ 
is the UV cutoff of the unified theory, which we expect to be within an 
order of magnitude from the reduced Planck scale $M_{\rm Pl}$.  As in 
the case of $H_u H_d$ in the MSSM, supergravity then generates an effective 
superpotential
\begin{equation}
  W_{\rm eff} = \frac{m_\Sigma}{2} \Sigma^2 
    + \frac{\lambda_\Sigma}{3} \Sigma^3,
\label{eq:Weff}
\end{equation}
where $m_\Sigma = c_2 m_{3/2}^*$ and $\lambda_\Sigma = c_3 
m_{3/2}^*/\Lambda$.  In this paper, we assume that the mediation scale 
of supersymmetry breaking $M_*$ is high, $M_* \sim \Lambda \sim M_{\rm Pl}$. 
This implies that $m_{3/2} = F_X/\sqrt{3}M_{\rm Pl} \sim \tilde{m} \approx 
O(F_X/M_*)$, where $F_X$ is the $F$-term vacuum expectation value (VEV) 
of supersymmetry breaking field $X$.  The key point is that the parameters 
appearing in Eq.~(\ref{eq:Weff}) are then $m_\Sigma \sim \tilde{m}$ and 
$\lambda_\Sigma \sim \tilde{m}/\Lambda$, which are both highly suppressed 
compared to conventional supersymmetric unified theories and are correlated 
with the scale of superpartner masses $\tilde{m}$.  An alternative way 
of obtaining the effective superpotential of Eq.~(\ref{eq:Weff}), which 
is available if the supersymmetry breaking field $X$ is singlet, is to 
have the K\"{a}hler potential terms
\begin{equation}
  K \,\supset\, \frac{c'_2}{2 M_*} X^\dagger \Sigma^2 
    + \frac{c'_3}{3 \Lambda M_*} X^\dagger \Sigma^3.
\label{eq:K'}
\end{equation}
In this case $m_\Sigma \sim \tilde{m}$ and $\lambda_\Sigma \sim 
\tilde{m}/\Lambda$, and these parameters are directly related with 
the superpartner mass scale $\tilde{m}$.

Assuming that the interactions in Eq.~(\ref{eq:Weff}) dominate over 
the soft supersymmetry breaking interactions for the scalar component 
$\phi_\Sigma$, which may not be an accurate approximation but is sufficient 
for the purpose of discussion here, we find a vacuum that breaks $SU(5)$ 
to $SU(3) \times SU(2) \times U(1)$
\begin{equation}
  \langle \Sigma \rangle 
  = \frac{V}{\sqrt{60}} \begin{pmatrix}
      2 & 0 & 0 & 0 & 0 \\  
      0 & 2 & 0 & 0 & 0 \\
      0 & 0 & 2 & 0 & 0 \\
      0 & 0 & 0 & -3 & 0 \\
      0 & 0 & 0 & 0 & -3
    \end{pmatrix},
\label{eq:Sigmavev}
\end{equation}
with $V = \sqrt{60} c_2 \Lambda /c_3$ (or $c_{2,3} \rightarrow c'_{2,3}$ 
if the relevant operators are as in Eq.~(\ref{eq:K'}), rather than 
Eq.~(\ref{eq:K})).  If $c_2/c_3 \sim 1$, the VEV of $\Sigma$ is so close 
to the UV cutoff that it is not reliably computed.  To avoid this we take 
$c_2$ to be sufficiently smaller than $c_3$, which is equivalent to taking 
$m_\Sigma$ to be sufficiently smaller than $\Lambda$ in conventional 
unified theories.

Adding soft supersymmetry breaking interactions $\tilde{m}_\Sigma^2 
\phi_\Sigma^\dagger \phi_\Sigma - \{ (B m_\Sigma/2) \phi_\Sigma^2 
+ (A \lambda_\Sigma/3) \phi_\Sigma^3 + {\rm h.c.} \}$, the precise 
locations of the minima of the potential will shift, leading to the 
naive expectation that $F_\Sigma \sim \tilde{m} \Lambda$.  Inserting 
this into the superpotential interaction $H \Sigma \bar{H}$ then yields 
$\mu B \sim \tilde{m} \Lambda$, so that a light Higgs doublet for 
electroweak symmetry breaking then requires $\mu  \sim \sqrt{\tilde{m} 
\Lambda} \gg \tilde{m}$.  However, such large supersymmetry breaking 
in the Higgs sector is forbidden because a negative mass-squared for 
the top squark arises at 1~loop order, leading to a very large spontaneous 
breaking of the color group.  The environmental requirement of an 
unbroken color symmetry forces a cancellation amongst terms so that 
$F_\Sigma \simlt 4\pi^2 \tilde{m}^2$.  This makes it probable that 
$\mu^2$ and $\mu B$ are an order of magnitude larger than $m^2_{H_{u,d}}$, 
leading to a Higgs quartic coupling of Eq.~(\ref{eq:lambdaexpansion}) 
with $\epsilon \sim 0.1$.

\section{Supersymmetry Breaking and Dark Matter}
\label{sec:multiverse}

Supersymmetry breaking is described by a spurion field $X$ that has a 
non-zero $F$ component, $F_X$.  This supersymmetry breaking is transmitted 
to the MSSM sector at a high mediation scale $M_* \sim M_{\rm Pl}$, so 
that $\tilde{m} = F_X/ M_*$ is not much larger than the gravitino mass 
$m_{3/2}$.  In general, we include all the supersymmetry breaking operators 
in the K\"{a}hler potential
\begin{align}
  {\cal L}_{\rm SB} &\sim \left[ 
    \left( \frac{X^\dagger X}{M_*^2} + \frac{X + {\rm h.c.}}{M_*} \right) 
      (T^\dagger T + \bar{F}^\dagger \bar{F} + H^\dagger H 
      + \bar{H}^\dagger \bar{H} + \Sigma^\dagger \Sigma) \right]_{\theta^4}
\nonumber\\
  &\quad {} + \left[ \left\{ \frac{X^\dagger}{M_*} (\Sigma^2 + \Sigma^3 
      + \cdots) + {\rm h.c.} \right\} + \frac{X^\dagger X}{M_*^2} 
      (\Sigma^2 + \Sigma^3 + \cdots) \right]_{\theta^4},
\label{eq:susybr}
\end{align}
where $\{ T, \bar{F} \}$ and $\{ H, \bar{H}, \Sigma \}$ are the matter 
and Higgs chiral superfields, and the terms associated with $H\bar{H}$ are 
assumed to be forbidden by the $R$ symmetry discussed below Eq.~(\ref{eq:W}). 
Note that each operator has an unknown coefficient of order unity that is 
not displayed.  We also allow the operator in the gauge kinetic function
\begin{equation}
  {\cal L}_{\rm SB}^\lambda \sim \left[ \frac{X}{M_*} 
    {\cal W}^\alpha {\cal W}_\alpha \right]_{\theta^2} + {\rm h.c.},
\label{eq:gauginomasses}
\end{equation}
where ${\cal W}^\alpha$ is the gauge field-strength superfield, but we 
forbid superpotential interactions coupling $X$ directly to matter and 
Higgs fields.  If $X$ is not neutral, some of the terms described above 
are forbidden.  For example, if it is charged under some symmetry, under 
which all the $SU(5)$ fields are neutral, then the terms linear in $X$ 
are all forbidden.

A significant model dependence for the superpartner spectrum may arise 
from the values of the Higgsino mass, $\mu$, and the gaugino masses 
$\tilde{m}_i$ for $i = SU(3), SU(2), U(1)$.  The $\mu$ parameter arises 
from the first two operators of Eq.~(\ref{eq:W}) and from supersymmetry 
breaking
\begin{equation}
  \mu = m_H - \frac{3}{\sqrt{60}}\, \lambda_H V + O(\tilde{m}).
\label{eq:mu}
\end{equation}
Fine-tuning between the first two terms is required so that $\mu$ is of order 
$\tilde{m}$ to ensure that the matrix of Eq.~(\ref{eq:higgsmasssquared}) 
has a small eigenvalue, allowing the scale of weak interactions to be far 
below $\tilde{m}$.  The gaugino masses, in general, have four contributions
\begin{equation}
  \tilde{m}_i = \tilde{m}_i^0 + \tilde{m}_i^{\rm AM} 
    + \tilde{m}_i^{\tilde{h}} + \tilde{m}_i^{\Sigma}.
\label{eq:gauginomasscontrib}
\end{equation}
The first term arises from Eq.~(\ref{eq:gauginomasses}) and occurs only 
if $X$ is neutral under all symmetries.  For $M_* > V$, this gives the 
contribution $\tilde{m}_i \approx \tilde{m} \, \alpha_i(\mu = \mu_c) / 
\alpha_5(\mu = M_*)$, where $\alpha_i = g_i^2/4\pi$ are the gauge field 
strength, with $\alpha_5$ being their unified value.  The remaining 
terms occur at one loop and are relevant only if $X$ carries a charge 
so that the leading term is absent.  The second term arises from anomaly 
mediation~\cite{Randall:1998uk} giving $\tilde{m}_i^{\rm AM} = b_i g_i^2\, 
m_{3/2} / 16\pi^2$, where $(b_1, b_2, b_3) = (33/5, 1 ,-3)$ are the 
beta-function coefficients, and we have taken the phase convention that 
$\tilde{m}_{1,2}$ are real and positive.  The third contribution is 
relevant only if the Higgsinos are heavier than the gauginos, as it 
arises from integrating out the Higgsinos, and contributes only to 
$\tilde{m}_{1,2}$.  The fourth term arises from integrating out the 
uneaten states of $\Sigma$, and contributes only to $\tilde{m}_{2,3}$.

We assume that $R$-parity is unbroken, so that the lightest supersymmetric 
particle (LSP) is stable and contributes to the dark matter if it is 
produced cosmologically.  We further assume environmental constraints 
that strongly disfavor observers in universes with much more dark matter 
than our own, as argued, e.g., in~\cite{Tegmark:2005dy,Bousso:2013rda}. 
This yields a large environmentally forbidden window in 
$m_{\rm LSP}$~\cite{Hall:2011jd}
\begin{equation}
  c_- {\rm TeV} < m_{\rm LSP} < c_+ T_R,
\label{eq:forbiddenLSPwindow}
\end{equation}
where $c_-$ depends on the LSP annihilation rate and is order unity 
while $c_+ \sim 10~\mbox{--}~30$, depending on the size of the reheating 
temperature after inflation, $T_R$.  If $m_{\rm LSP} > c_- {\rm TeV}$ 
the LSP is overproduced at freezeout, unless  $m_{\rm LSP} > c_+ T_R$ 
so that it never gets close to thermal equilibrium.  Taking $\tilde{m}$ 
to be fixed around $\mu_c \simeq 10^{10}~\mbox{--}~10^{12}~{\rm GeV}$, 
as determined by the $125~{\rm GeV}$ Higgs boson mass, it is critical 
whether the resulting value of $m_{\rm LSP}$ is above or below $c_+ T_R$.
\begin{itemize}
\item If $m_{\rm LSP} < c_+ T_R$, fine-tuning between contributions to 
$m_{\rm LSP}$ is required to force $m_{\rm LSP} < c_- {\rm TeV}$.  Dark 
matter will then have a sizable LSP component, although there could be 
other sizable components such as axions.
\item If $m_{\rm LSP} \geq c_+ T_R$, there are two possibilities.  If there 
is non-LSP dark matter, such as axions, then an environmental selection 
for dark matter does not force any fine-tuning in the superpartner masses 
(beyond that already required in $\mu$ to allow electroweak symmetry 
breaking).  The LSP contribution to dark matter will be negligible, and 
no superpartners are expected in the TeV domain.  On the other hand, if 
there is no axion or other source for dark matter, then an environmental 
requirement for significant dark matter will force $m_{\rm LSP} \sim 
c_+ T_R$, so that dark matter is produced at reheating from collisions 
involving the high energy tail of thermal distributions.  In this case, 
the LSP will contribute all of the dark matter.  No superpartners, however, 
are expected in the TeV domain, since the LSP is rather heavy of mass 
$\sim c_+ T_R$.
\end{itemize}
Which of these scenarios is realized is model dependent.  Below 
we consider three representative cases.  We first consider in 
Section~\ref{subsec:pure} a class of models that requires no fine-tuning 
of $m_{\rm LSP}$: the first case discussed in the second bullet point 
given above.  In this case, the LSP is extremely heavy and its nature 
is rather unimportant.  We then consider two models of the type described 
in the first bullet point, in which fine-tuning of the LSP mass is 
required.  Note that whenever the LSP mass is environmentally forced 
to be much less than its typical value, the LSP will be a fermion, since 
the tuning in $m_{\rm LSP}$ is linear for a fermion and quadratic for 
a boson.  Hence in one of these models the LSP is a wino and in the 
other a Higgsino, which we will discuss in Sections~\ref{subsec:wino} 
and \ref{subsec:higgino}, respectively.

\subsection{Pure Intermediate Scale Supersymmetry}
\label{subsec:pure}

A particularly simple possibility is that all the superpartner masses are 
within one or two orders of magnitude of the scale $\tilde{m}$ determined 
by Eq.~(\ref{eq:quartic-bc}).  This would occur, for example, if all 
the supersymmetry breaking operators in Eqs.~(\ref{eq:susybr}) and 
(\ref{eq:gauginomasses}) occur with unsuppressed coefficients.  In 
this case we take the reheating temperature after inflation to be 
sufficiently low that none of the superpartners have a significant 
cosmological abundance.  This requires baryogenesis at a relatively 
low temperature which could result, for larger values of $\tilde{m}$, 
from thermal leptogenesis.  Dark matter would be non-supersymmetric, 
for example axions.

\begin{figure}[t]
\begin{center}
  \subfigure{\includegraphics[height=5.8cm]{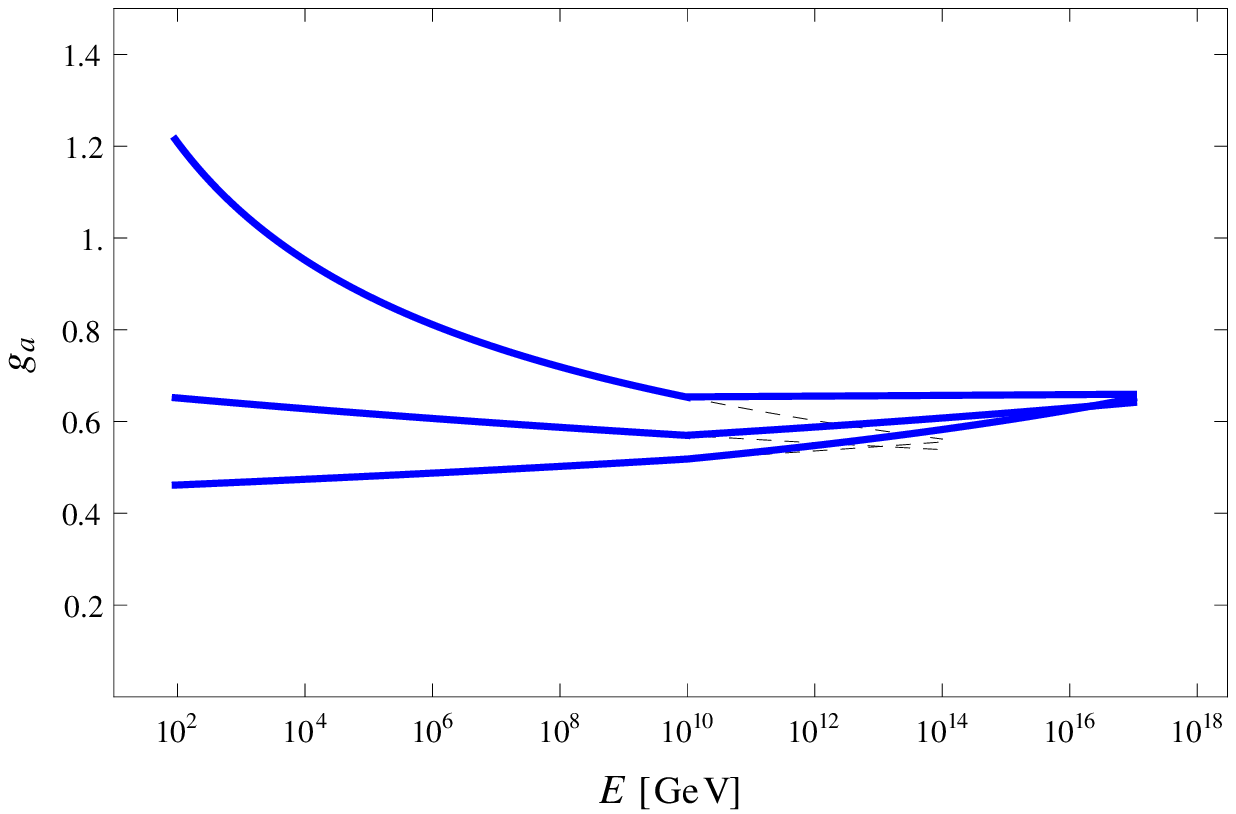}}
\hspace{0.1cm}
  \subfigure{\includegraphics[height=5.8cm]{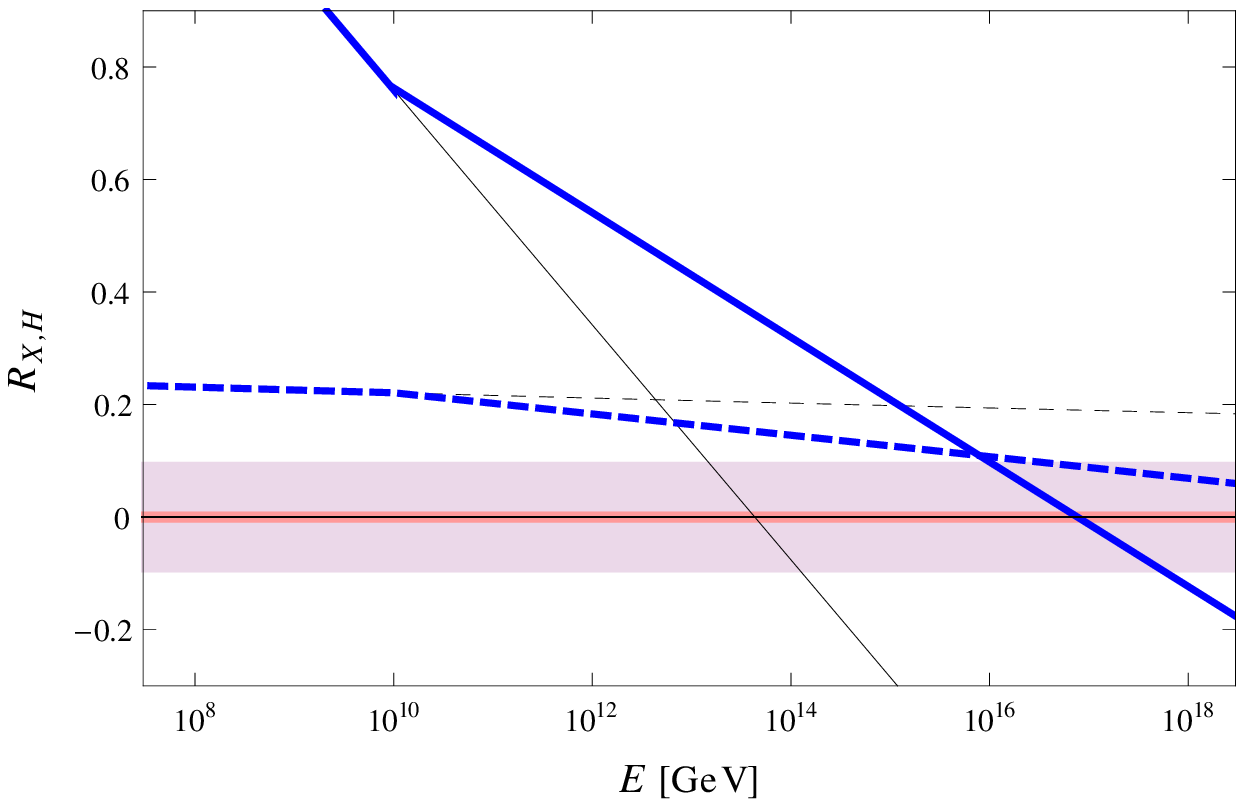}}
\caption{{\bf Pure Intermediate Scale Supersymmetry Breaking.} Left panel:\ 
 evolution of gauge couplings in ISS (solid, blue) and in the SM (dashed, 
 black).  Right panel:\ evolution of $R_X$ (solid) and $R_H$ (dashed) for 
 ISS (thick, blue) and the SM (thin, black).  Without corrections from 
 dimension~5 and 6 operators, $M_X$ and $M_H$ are the scales at which 
 $R_X$ and $R_H$ vanish, respectively.  Including the dimension~5 operator, 
 allowed values of $M_H$ occur when the dashed thick (blue) line lies 
 inside the pale (red) shaded region.  Including the dimension~6 operators, 
 allowed values of $M_X$ occur when the solid thick (blue) line lies 
 inside the dark (red) shaded region.  Both shaded (red) bands have been 
 drawn for $V/\Lambda = 0.1$.}
\label{fig:Pure}
\end{center}
\end{figure}

The unification of gauge couplings for this class of models is illustrated 
by the solid blue lines in the left panel of Fig.~\ref{fig:Pure}.  The 
masses of all superpartners and uneaten components of $\Sigma$ have a 
reference value of $M_I = 10^{10}~{\rm GeV}$ in all figures.   The dashed 
lines show the corresponding curves for the SM.  The unification now 
occurs above $10^{16}~{\rm GeV}$, rather than near $10^{14}~{\rm GeV}$ 
as in the SM.  Running is computed at two-loop order.

To understand the nature of the unification more precisely, it is useful 
to look at the running of two particular combinations of the gauge 
couplings~\cite{Hisano:1992mh}.  One combination
\begin{equation}
  R_X =  \frac{1}{\sqrt{38}} \left( \frac{5}{g_1^2} 
    - \frac{3}{g_2^2} - \frac{2}{g_3^2} \right),
\label{eq:RX}
\end{equation}
is shown as the solid line in the right panel of Fig.~\ref{fig:Pure}, thick 
blue for ISS and thin black for the SM.  The other combination
\begin{equation}
  R_H = \frac{1}{\sqrt{14}} \left( \frac{3}{g_2^2} 
    - \frac{2}{g_3^2} - \frac{1}{g_1^2} \right),
\label{eq:RH}
\end{equation}
is shown by the dashed lines.  These are key combinations since in the 
minimal $SU(5)$ model with ISS, $R_{X,H}$ vanish at energies $M_{X,H}$, 
the masses of the heavy $SU(5)$ gauge bosons and the heavy color triplets 
that are the $SU(5)$ partners of the Higgs doublets, respectively.  Hence 
$M_X$ and $M_H$ for the reference values of the low energy particle masses 
can be read off directly from the right panel of Fig.~\ref{fig:Pure} as 
the energy at which the solid and dashed lines cross the horizontal axis.

The deficiencies of gauge coupling unification in the SM are emphasized 
in this figure---$M_X$ is two orders of magnitude below the current limit 
from proton decay, whereas $M_H$ is very far above the Planck scale. 
While the latter can be cured by a dimension-5 operator of the form 
$[\Sigma {\cal W}^\alpha {\cal W}_\alpha]_{\theta^2}$, additional physics 
is required to satisfy the experimental constraint from proton decay.  With 
ISS, it is the combination of the gauginos, the color octet, $\Sigma_8$, 
and the weak triplet, $\Sigma_3$ that increase the $SU(5)$ gauge boson 
mass to $M_X \simeq 7.3 \times 10^{16}~{\rm GeV}$ at the reference point 
of $m_\lambda = M_{\Sigma_8} = M_{\Sigma_3} = 10^{10}~{\rm GeV}$.  The 
reference point value for the colored Higgs mass is $M_H \simeq 5.5 \times 
10^{21}~{\rm GeV}$, still above the Planck scale, but needs only about 
half the correction of the SM to reduce it to of order $V \sim M_X$.

It is straightforward to include the effect of deviating from the reference 
spectrum.  We then find
\begin{align}
  \ln \left( \frac{M_X}{7.3 \times 10^{16}~{\rm GeV}} \right) 
  &= -\frac{1}{6} \ln \frac{m_{\tilde{g}} m_{\tilde{w}}}{M_I^2} 
    - \frac{1}{4} \ln \frac{m_{\Sigma_8} m_{\Sigma_3}}{M_I^2} 
    + \delta_X^I + \delta_X^U,
\label{eq:MX}\\
  \ln \left( \frac{M_H}{5.5 \times 10^{21}~{\rm GeV}} \right) 
  &= \frac{1}{6} \ln \frac{m^4_{\tilde{h}} m_H}{M_I^5} 
    + \delta_H^I + \delta_H^U,
\label{eq:MH}
\end{align}
where
\begin{equation}
  \delta_X^I = -\frac{1}{8} 
    \ln \frac{m^2_{\tilde{q}}}{m_{\tilde{u}}m_{\tilde{e}}},
\qquad
  \delta_H^I= -\frac{5}{3} \ln \frac{m_{\tilde{g}}}{m_{\tilde{w}}} 
    - \frac{5}{2} \ln\frac{m_{\Sigma_8}}{m_{\Sigma_3}} 
    + \frac{1}{4} \ln \frac{m_{\tilde{q}}^4}{m_{\tilde{u}}^3 m_{\tilde{e}}} 
    - \frac{1}{2} \ln \frac{m_{\tilde{d}}}{m_{\tilde{l}}},
\label{eq:deltaXH}
\end{equation}
parameterize squark/slepton and $\Sigma_{8/3}$ non-degeneracies, and the 
corrections $\delta_{X,H}^U$ contain possible terms from non-degenerate 
unified multiplets in non-minimal models.  Here, the deviations of $M_{X,H}$ 
from the reference values have been calculated at the 1-loop order.  We 
find that a deviation of the low energy spectrum from the reference mass 
of $10^{10}~{\rm GeV}$ does not significantly affect the values of $M_X$ 
or $M_H$, unless a large deviation is considered.  In particular, it is 
hard to bring $M_H$ down to of order $M_X$ to make the unification work 
by this effect alone.

The required threshold correction, however, easily arises from 
higher-dimensional gauge kinetic operators.  At the leading order, we 
may consider the dimension-5 operator of the form $[\Sigma {\cal W}^\alpha 
{\cal W}_\alpha]_{\theta^2}$, which affects $M_H$ significantly and can 
easily bring it down to $\sim 10^{17}~{\rm GeV}$.  Interestingly, this 
operator does not affect $M_X$, so that the prediction for the heavy gauge 
boson mass stays as in Eq.~(\ref{eq:MX}).  The correction to $M_X$ appears 
only at the next order from the operator of the form $[(\Sigma {\cal 
W}^\alpha) (\Sigma {\cal W}_\alpha)]_{\theta^2}$.  To see these explicitly, 
we may replace the canonical gauge kinetic term as
\begin{equation}
  {\rm Tr}[{\cal W}^\alpha {\cal W}_\alpha] 
  \rightarrow {\rm Tr}[{\cal W}^\alpha {\cal W}_\alpha] 
    + \frac{a}{\Lambda} {\rm Tr}[\langle \Sigma \rangle 
      {\cal W}^\alpha {\cal W}_\alpha] 
    + \frac{b}{\Lambda^2} {\rm Tr}[\langle \Sigma \rangle {\cal W}^\alpha]\, 
      {\rm Tr}[\langle \Sigma \rangle {\cal W}_\alpha],
\label{eq:higherdimkinetic}
\end{equation}
where $\Lambda$ is the UV cutoff of the unified theory.  We then find 
that the right-hand sides of Eqs.~(\ref{eq:MX}) and (\ref{eq:MH}) receive 
corrections of $(5\pi^2 b/6 g_U^2) (V^2/\Lambda^2)$ and $-(20\pi^2 a/\sqrt{15} 
g_U^2) (V/\Lambda) + O(V^2/\Lambda^2)$, respectively, where $g_U$ is the 
unified gauge coupling at the scale $V$.  In terms of $R_X$ and $R_H$, 
these correspond to
\begin{equation}
  \varDelta R_X = \frac{5\, b}{2\sqrt{38}\, g_U^2} \frac{V^2}{\Lambda^2} 
  \simeq 0.96\, b\, \frac{V^2}{\Lambda^2},
\qquad
  \varDelta R_H = -\frac{\sqrt{6}\, a}{\sqrt{35}\, g_U^2} \frac{V}{\Lambda} 
  \simeq -0.98\, a\, \frac{V}{\Lambda}.
\label{eq:DeltaR}
\end{equation}
Namely, if the solid and dashed lines in the right panel of 
Fig.~\ref{fig:Pure} enter in the ranges of $\pm \varDelta R_X$ and 
$\pm \varDelta R_H$, the running of the gauge couplings can be made 
consistent with the $SU(5)$ unification.

In the right panel of Fig.~\ref{fig:Pure}, the corrections for $R_X$ and 
$R_H$ in Eq.~(\ref{eq:DeltaR}) are depicted by the dark (red) and light 
(pink) shaded regions, respectively, for $V/\Lambda =  0.1$ with $a \in 
[-1,1]$ and $b \in [-1,1]$.  While the second order correction $\varDelta 
R_X$ is small, the first order correction $\varDelta R_H$ is about $4\%$ 
and is sufficient to lower $M_H$ so that it is comparable to $M_X$. 
Since this can be done with a sufficiently small value of $V/\Lambda$, 
unification works well in this class of models.

With the values of $M_{X,H}$ and $g_U$ obtained above, the rate for proton 
decay is expected to be small.  Since the leading correction from the 
dimension-5 operator in Eq.~(\ref{eq:higherdimkinetic}) does not affect 
$M_X$, it is a relatively robust result that proton decay due to heavy 
gauge boson exchange is suppressed in the pure ISS models described here. 
However, it is possible that the rate is near the current experimental 
bound and within reach of future searches if the superparticle mass scale 
is relatively high, $\tilde{m} \sim 10^{11}~{\rm GeV}$, because of the 
uncertainty of $M_X$ from the spectrum of intermediate scale particles; 
see Eq.~(\ref{eq:MX}).

\subsection{ISS with TeV-scale Wino}  
\label{subsec:wino}

If the values of $m_{\rm LSP}$ allowed by scanning are less than $T_R$, 
environmental selection is likely to force a large amount of fine-tuning 
in $m_{\rm LSP}$, so that the amount of dark matter arising from LSP 
freezeout is limited.  Possible selection effects include close stellar 
encounters, galactic disk stability~\cite{Tegmark:2005dy} and baryon 
dilution~\cite{Bousso:2013rda}.  These selection effects act on any dark 
matter candidate, no matter the production mechanism, so dark matter may 
be multi-component, with the contribution from each component being the 
same order of magnitude.  If the LSP is a gaugino then it must be the 
wino; a light bino will not annihilate efficiently and a light gluino 
as a significant component of dark matter is excluded.  A light wino can 
arise from a cancellation among the various contributions to $\tilde{m}_2$ 
in Eq.~(\ref{eq:gauginomasscontrib}).  If $X$ is neutral the coefficient 
of the $[X {\cal W}^\alpha {\cal W}_\alpha]_{\theta^2}$ interaction 
must be small to allow such a cancellation; while if $X$ is charged, so 
$\tilde{m}_2^0 = 0$, the cancellation occurs among the loop contributions 
in Eq.~(\ref{eq:gauginomasscontrib}).

\begin{figure}[t]
\begin{center}
  \subfigure{\includegraphics[height=5.8cm]{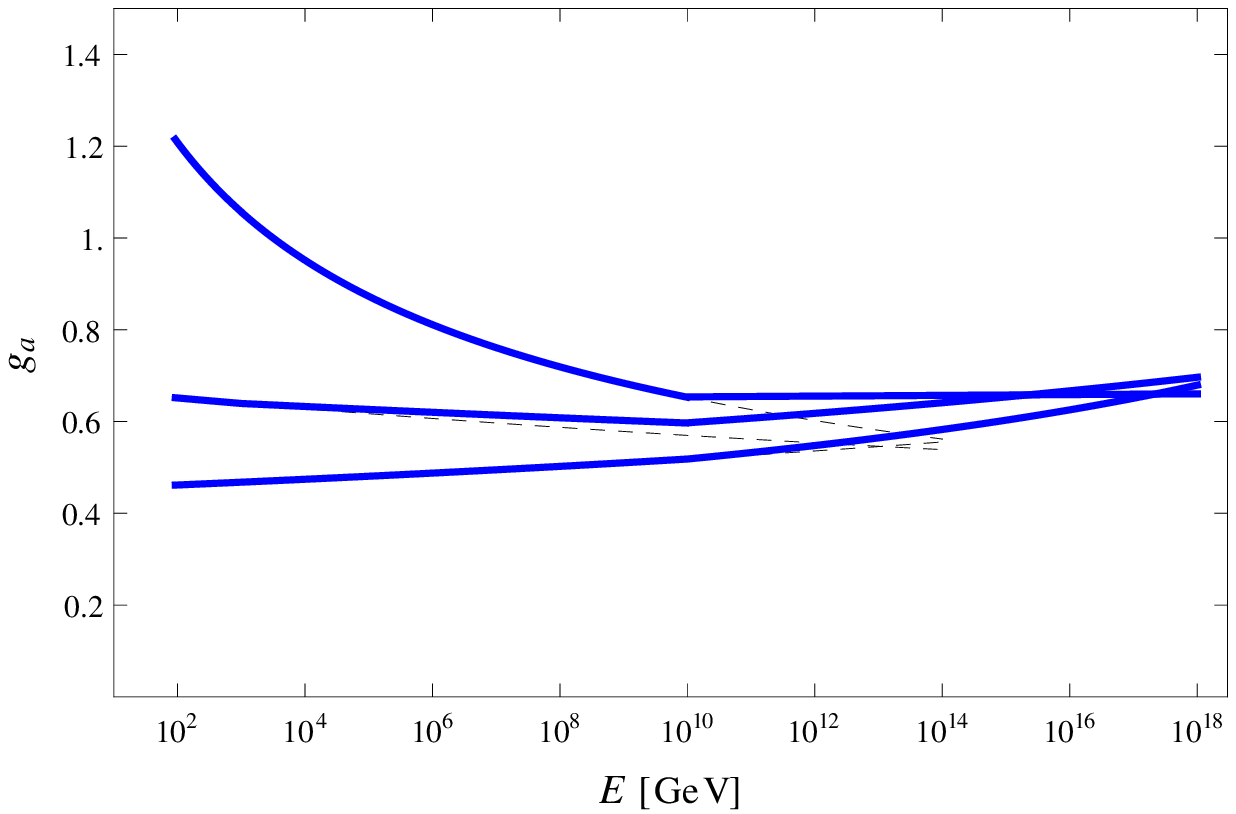}}
\hspace{0.1cm}
  \subfigure{\includegraphics[height=5.8cm]{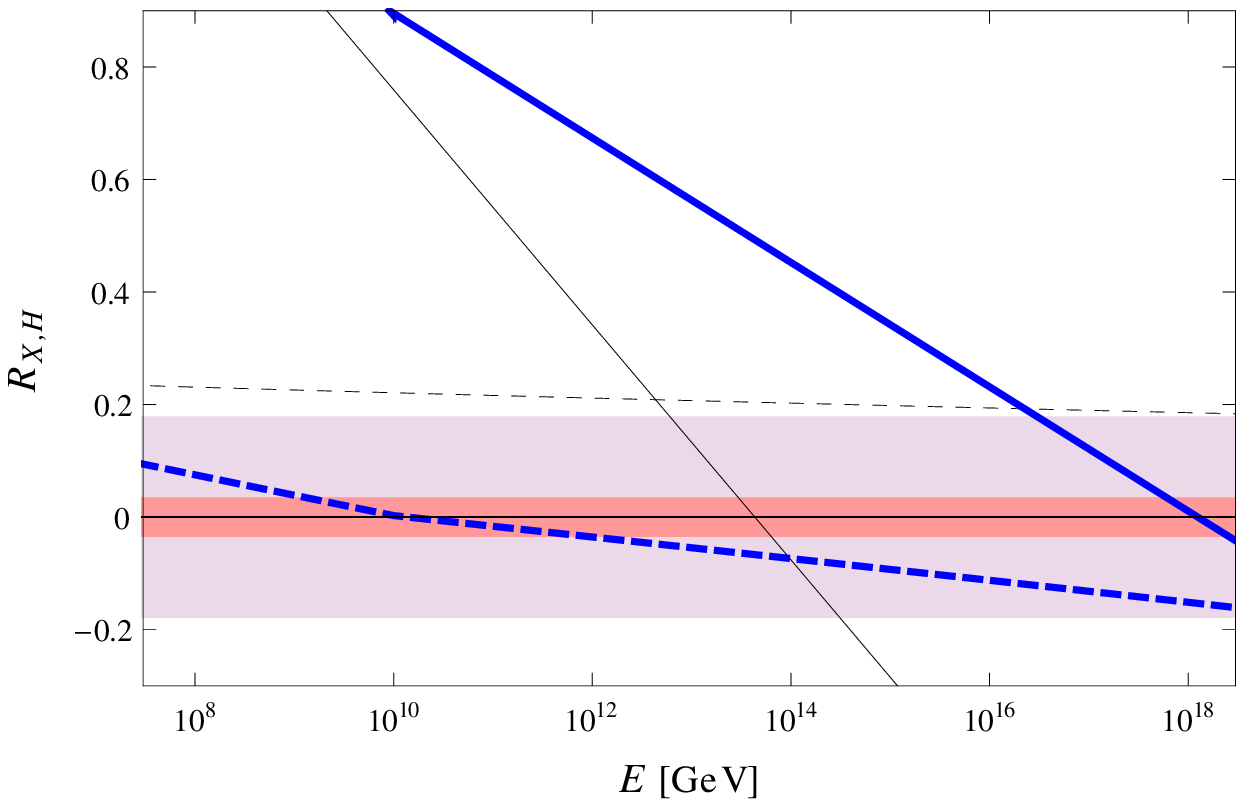}}
\caption{{\bf Intermediate Scale Supersymmetry Breaking with TeV-scale wino.} 
 Caption as in Fig.~\ref{fig:Pure}, except both shaded (red) bands have been 
 drawn for $V/\Lambda = 0.2$.}
\label{fig:Wino}
\end{center}
\end{figure}

Gauge coupling unification for ISS with the TeV-scale wino is shown 
in Fig.~\ref{fig:Wino}.  Lowering the wino mass pushes up $M_X$ by 
over an order of magnitude since, from Eq.~(\ref{eq:MX}), $M_X \sim 
1/m_{\tilde{w}}^{1/6}$, so that proton decay from $X$ exchange is greatly 
suppressed.  On the other hand, from Eq.~(\ref{eq:MH}), $M_H \sim 
m_{\tilde{w}}^{5/3}$ so $M_H$ is reduced all the way to the intermediate 
scale.  A correction from the dimension-5 operator $[\Sigma {\cal W}^\alpha 
{\cal W}_\alpha]_{\theta^2}$ in Eq.~(\ref{eq:higherdimkinetic}), however, 
can raise it back to $\sim M_X$, avoiding excessive proton decay 
from the colored Higgs triplet exchange.   For $M_H$ to be near $M_X$ 
requires $a V/ \Lambda \sim -0.2$, as shown by the shading in the right 
panel.  Since $M_X$ is so large it is reasonable that the interval of 
validity of the unified theory is restricted.

The H.E.S.S. Collaboration is currently probing wino dark matter indirectly, 
via searches for monochromatic gamma lines~\cite{Abramowski:2013ax}.  For 
a NFW halo profile, wino dark matter is excluded from comprising more than 
about $20\%$ of dark matter~\cite{Cohen:2013ama}.  However, the uncertainties 
from the halo profile are large, and our environmental scheme makes 
multi-component dark matter plausible so that the wino comprises only 
a part of the dark matter.

\subsection{ISS with TeV-scale Higgsino}  
\label{subsec:higgino}

The remaining possibility for $m_{\rm LSP} \leq T_R$ is for the Higgsino 
mass environmentally tuned to be in the TeV range.  The Higgsino mass 
parameter, $\mu$ given in Eq.~(\ref{eq:mu}), undergoes tuning to the 
intermediate scale to allow electroweak symmetry breaking, and further 
tuning to the TeV scale allows a substantial Higgsino dark matter component. 
In the case that the wino and bino masses are of order $10^{10}~{\rm GeV}$ 
this is excluded from direct detection probes of galactic dark matter, 
since in these experiments the Higgsinos act as a Dirac particle with 
couplings to the $Z$ boson.  However, if the supersymmetry breaking 
field $X$ is charged, so that $\tilde{m}^0_i = 0$, the remaining 
loop-induced gaugino masses have just the right size to render 
the Higgsinos as inelastic dark matter in these direct detection 
experiments~\cite{TuckerSmith:2001hy}, with a mass splitting between 
the two neutral Majorana Higgsinos of order $200~{\rm keV}$.

\begin{figure}[t]
\begin{center}
  \subfigure{\includegraphics[height=5.8cm]{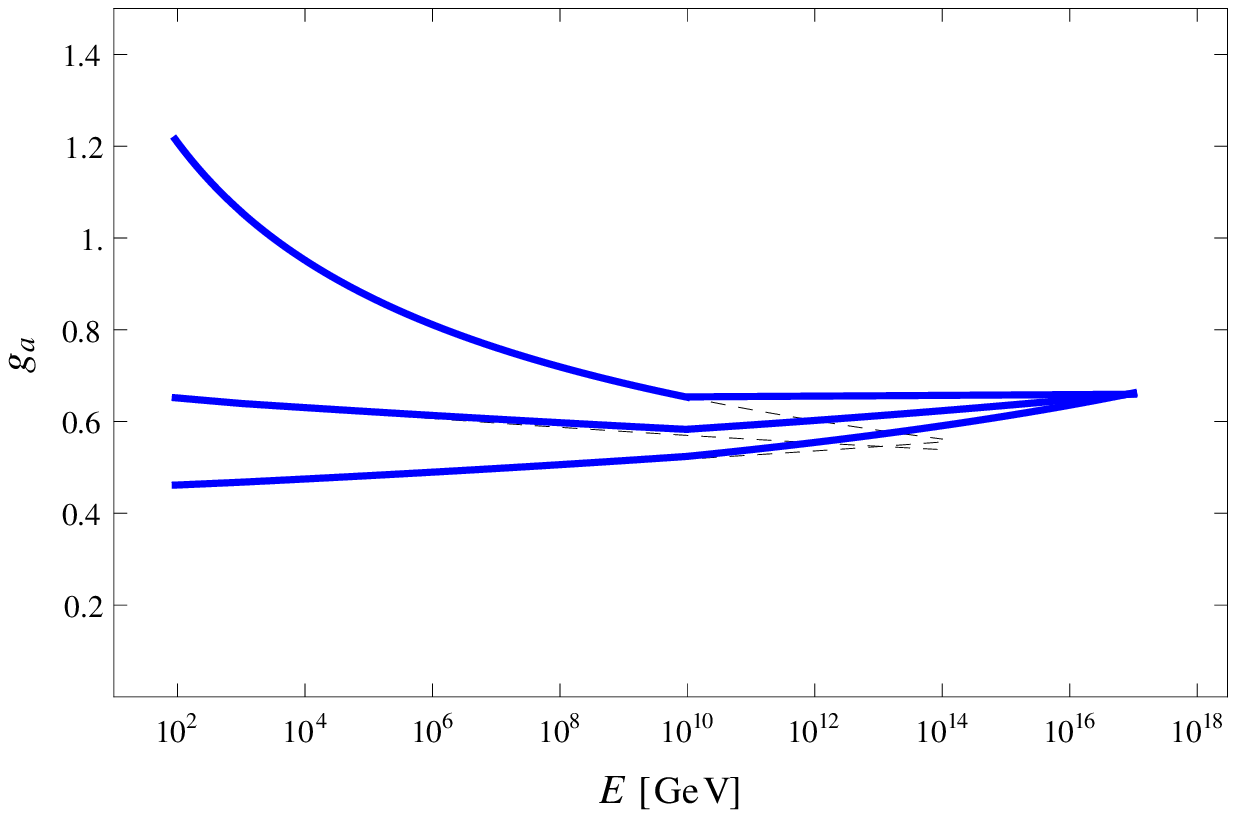}}
\hspace{0.1cm}
  \subfigure{\includegraphics[height=5.8cm]{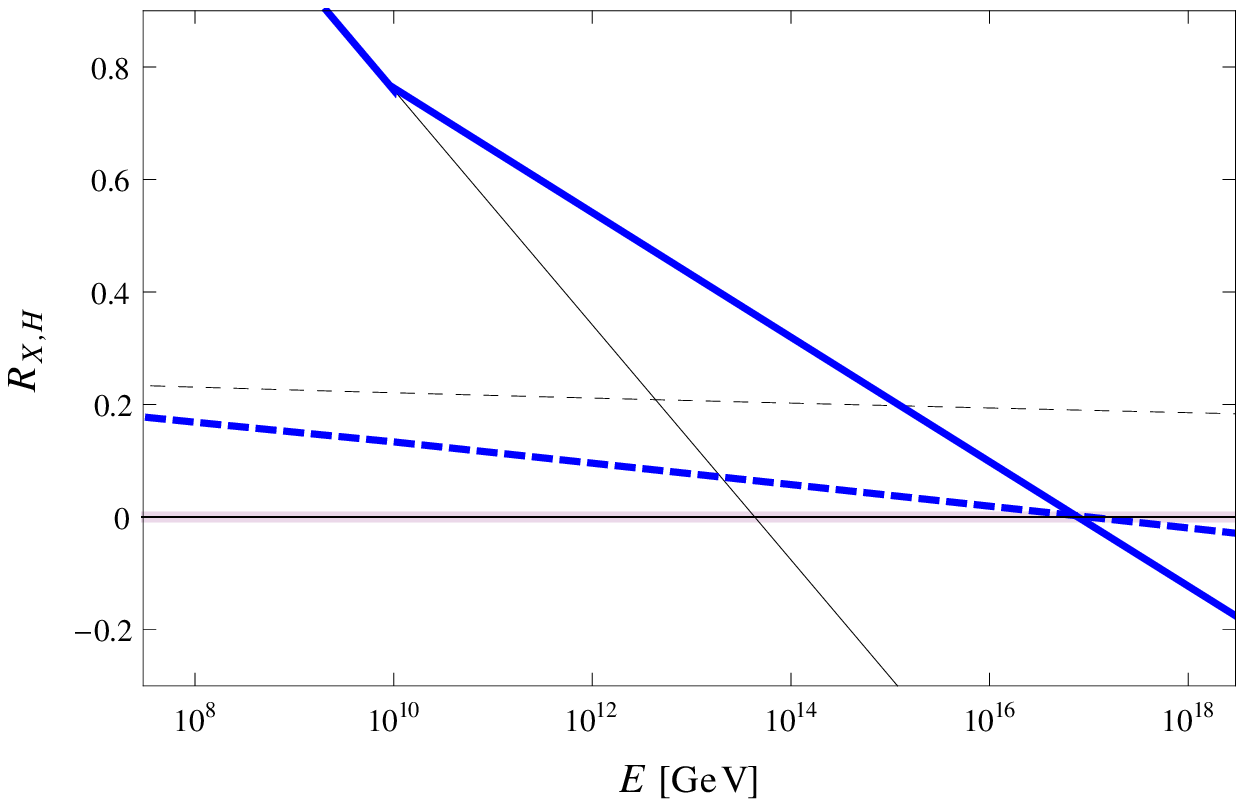}}
\caption{{\bf Intermediate Scale Supersymmetry Breaking with TeV-scale 
 Higgsino.}  Caption as in Fig.~\ref{fig:Pure}, except both shaded (red) 
 bands have been drawn for $V/\Lambda = 0.01$.}
\label{fig:Higgsino}
\end{center}
\end{figure}

Gauge coupling unification for ISS with the TeV-scale Higgsino is shown 
in Fig.~\ref{fig:Higgsino}.  Lowering the Higgsino mass does not alter 
$M_X$, so the prediction for proton decay from $X$ exchange is essentially 
as in the case of pure ISS discussed in Section~\ref{subsec:pure}.  On 
the other hand, since $M_H \sim m_{\tilde{h}}^{2/3}$, $M_H$ is lowered to 
be very close to $M_X$ without a threshold correction from higher-dimension 
operators in Eq.~(\ref{eq:higherdimkinetic}).  The accuracy of this 
unification is comparable to the MSSM so that, with $|a| \sim 1$, the 
cutoff scale $\Lambda$ could be two orders of magnitude larger than 
the $SU(5)$-breaking VEV $V$.

\section*{Acknowledgments}

We recently learned that Patrick Fox, Graham Kribs, and Adam Martin are 
preparing a paper on Dirac gauginos where the scale of supersymmetry 
breaking is linked to the scale at which the SM Higgs quartic vanishes. 
We thank Graham Kribs for useful communications and discussions.  We also 
thank Satoshi Shirai for discussions.  This work was supported in part by 
the Director, Office of Science, Office of High Energy and Nuclear Physics, 
of the US Department of Energy under Contract DE-AC02-05CH11231 and in 
part by the National Science Foundation under grants PHY-0855653 and 
PHY-1214644.

\end{document}